\documentclass[conference]{IEEEtran}
\IEEEoverridecommandlockouts
\usepackage{cite}
\usepackage{amsmath,amssymb,amsfonts}
\usepackage{algorithmic}
\usepackage{graphicx}
\usepackage{textcomp}
\usepackage{xcolor}
\usepackage{booktabs}
\usepackage{array}
\usepackage{multirow}
\usepackage{enumitem}
\usepackage{url}
\usepackage{tikz}
\usepackage{pifont}
\usepackage{hyperref}
\def\BibTeX{{\rm B\kern-.05em{\sc i\kern-.025em b}\kern-.08em
    T\kern-.1667em\lower.7ex\hbox{E}\kern-.125emX}}
\begin{document}

\title{AROhI: An Interactive Tool for Estimating ROI of Data Analytics}

\author{
\IEEEauthorblockN{Noopur Zambare\IEEEauthorrefmark{1},
Jacob Idoko\IEEEauthorrefmark{2},
Jagrit Acharya\IEEEauthorrefmark{2},and
Gouri Ginde\IEEEauthorrefmark{2}}
\IEEEauthorblockA{\IEEEauthorrefmark{1}\textit{Dept. of Mechanical Engineering, Indian Institute of Technology Jodhpur, India}\\
Email: \ zambare.1@iitj.ac.in}
\IEEEauthorblockA{\IEEEauthorrefmark{2}\textit{Dept. of Electrical and Software Engineering, University of Calgary, Canada}\\
Email: \{jacob.idoko, jagrit.acharya1, gouri.deshpande\}@ucalgary.ca}
}

\maketitle

\begin{abstract}
The cost of adopting new technology is rarely analyzed and discussed, while it is vital for many software companies worldwide. Thus, it is crucial to consider Return On Investment (ROI) when performing data analytics. Decisions on "How much analytics is needed"? are hard to answer. ROI could guide decision support on the What?, How?, and How Much? Analytics for a given problem. This work details a comprehensive tool that provides conventional and advanced ML approaches for demonstration using requirements dependency extraction and their ROI analysis as use case. Utilizing advanced ML techniques
such as Active Learning, Transfer Learning and primitive Large language model: BERT (Bidirectional Encoder Representations from Transformers) as its various components for automating dependency extraction, the tool outcomes demonstrate a mechanism to compute the ROI of ML algorithms to present a clear picture of trade-offs between the cost and benefits of a technology investment.
\end{abstract}


\section{Introduction}
Most Machine Learning (ML) solutions are evaluated based on the algorithm's performance measures, such as accuracy and AUC-ROC (Area Under the Receiver Operating Characteristic) Curve, precision and recall. However, ML algorithms for a particular problem typically consume more computational resources than classical algorithms for the same problem. Furthermore, these methods require large training (annotated) datasets, which tends to become an effort and cost-intensive aspect. Hence, decisions on "How much is needed?" are hard to answer; however, ROI could be used as guidance for decision support in deciding What? How? and How Much? of analytics for a given problem\cite{deshpande2020beyond}.

Our contributions to this work are as follows:
\begin{itemize}
\item Provide a method to compare and evaluate the ML approaches not only using ML performance measures but also based on ROI criteria (based on our prior research work \cite{deshpande2020beyond}, \cite{deshpande2021bert})
    \item Propose a comprehensive interactive tool design to visualize the ROI and ML performance measurement trade-offs.
    \item Demonstrate how users can change cost factors and view the projections of changing ROI for decision-making through a dashboard-like interactive interface.  
    \item Demonstrate the utility of the solution application through a requirements dependency extraction use case.
    \item Developed and hosted AROhI  tool\footnote{Excluding URL to tool for double-blind review} on the AWS instance for everyone to use.
    
\end{itemize}
\section{Related Work}
In recent years, the assessment of the ROI for ML solutions across various domains has increased. This section explores existing research, tools, and frameworks, focusing on ROI estimation in data analytics, particularly for ML and AI applications.

\subsection{Existing frameworks for measuring the business value of Data Analytics}

Several frameworks have been developed to enhance ROI estimation within data analytics and ML projects. These tools provide methodologies to estimate AI-driven solutions' financial impact and value. Some notable examples among them include:

\begin{itemize}
    \item Net Present Value (NPV): NPV calculates the present value of a data analytics project's expected future cash flows. It considers the initial investment, expected returns, and the time value of money \cite{brigham2016financial}\cite{koller2010valuation}.

    \item Cost-Benefit Analysis: CBA compares the costs associated with implementing and maintaining a data analytics project with the benefits it generates. It considers tangible and intangible costs and benefits to determine the overall ROI \cite {baracskay1998cost}.

    \item Return on Investment Ratio: A simple ratio that compares the net benefits of a data analytics project to its costs. It is expressed as a percentage and provides a quick snapshot of the project's financial performance \cite {phillips2012return}.

    \item Break-Even Analysis: This model determines the point at which the benefits of a data analytics project equal its costs. It helps identify the time it takes for the project to reach a break-even point and start generating positive returns \cite{siegel1998schaum}.

    \item Value-Based ROI: This framework focuses on measuring the value created by a data analytics project in terms of its impact on key performance indicators (KPIs) or strategic objectives. It goes beyond financial metrics to assess the project's overall value proposition \cite{stewart1991quest}.

    \item Total Cost of Ownership (TCO): TCO considers the total lifecycle costs of a data analytics project, including upfront investment, operational costs, maintenance, and potential upgrades. It provides a comprehensive view of the project's financial impact \cite{ellram1995total}.
    
\end{itemize}

\subsection{Industry applications of ROI}
ROI estimation tools and frameworks are commonly used in various sectors to improve decision-making processes and boost business outcomes with the help of data analysis and ML. Some specific uses in the industry include:

\begin{itemize}
    \item Finance: ROI estimating tools and frameworks play an important role in finance by optimizing investment strategies and predicting trends in the market. Lee et al. (2022) show that AI models such as LSTM improve decision-making in real estate investments and reveal high ROI \cite{lee2022prediction}. Net Present Value (NPV) and ROI ratio are used to measure AI/ML-based investment strategies' financial feasibility and performance \cite{npvFinance}.

    \item Healthcare: Healthcare is a field that AI and data analytics have greatly influenced in enhancing patients' quality of life. Pandey et al. (2021) highlight AI's impact on enhancing ROI by improving the efficiency of operations while at the same time reducing the costs involved \cite{pandey2021roi}. Cost Benefit Analysis (CBA) analyzes the economic benefits of healthcare performance indicators such as patient recovery rates and the use of resources \cite{brent2023cost}.
    
    \item Manufacturing and Aerospace: AI and robotics integration in manufacturing and aerospace increases productivity and operational efficiency, thereby delivering rapid ROI. Marjanović et al. (2018) show that integrating value and sustainability assessments in design space exploration through ML optimizes design processes and reduces costs \cite{bertoni2018model}. On the other hand, Figy (2021) discusses how Labplas achieved an ROI of less than one year by redesigning custom production machines with EtherCAT, robotics, and AI for quality control. This resulted in a 15 to 35\% increase in machine productivity \cite{figy2021pc}. Break-even Analysis is used to identify the point at which the benefits of the new technology equal its costs. At the same time, Total Cost Ownership (TCO) evaluates the full lifecycle costs against the gains in productivity.
\end{itemize}
\textbf{However, to our knowledge,} no tool has been developed yet that can be used to determine the value of data analytics for any given problem through a no-code interaction interface. This study is a step towards achieving such a feat.

\begin{figure*}[!t]
    \centering
     \includegraphics[scale=.24]{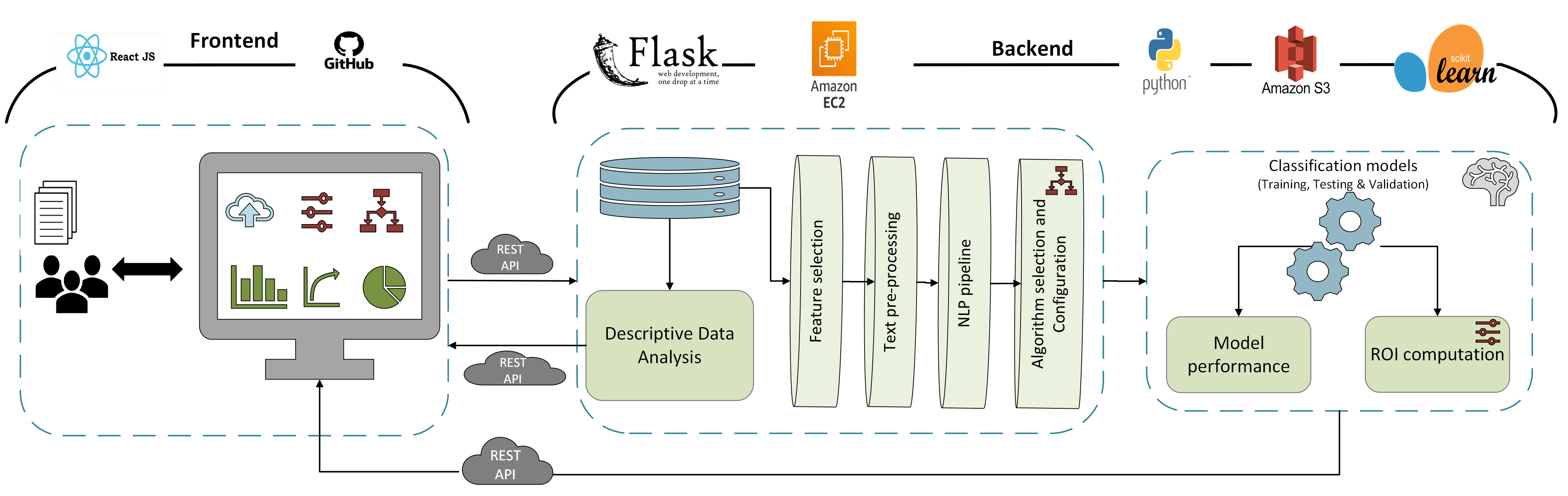}
    \caption{Conceptual design of the AROhI tool and various technology components of it }
    \label{fig: Web Tool Framework}
\end{figure*}

\section{Need for ROI estimation}
\subsection{The accuracy fallacy}
In most ML projects, accuracy is often the most reported metric for evaluating the performance of a model. However, this can be misleading and result in a fallacy that undermines the actual value of a model. The "accuracy fallacy" here is that a model's high accuracy equates to reliability and performance, which is rarely true. Accuracy measures how often a model's predictions are correct. However, it does not account for the cost of errors, such as false positives and false negatives. For instance, a model that predicts rare events can achieve high accuracy by simply predicting the majority class, which does not provide meaningful insight into the minority class. 
Accuracy also fails to account for the balance between recall (the ability to find all relevant instances) and precision (the ability to find relevant instances) \cite{dimatteo2020requirements}. In most cases, recall should carry more weight than precision, as noted by Berry et al. in their empirical evaluation of tools for requirements engineering tasks \cite{8049186}. These limitations call for more robust evaluation metrics that align with business objectives.

To truly evaluate a model's effectiveness, it is necessary to estimate its ROI. This involves considering factors such as lift, which measures how much better the model is compared to random guessing, and the specific costs associated with different types of errors \cite{siegel2024ai}. In summary, it captures the multiplicative improvement to operations and the increased return for your efforts, i.e., ROI. Wafiq Syed, a Data Product Manager at Walmart, notes, "The most important metric for your model's performance is the business metric that it is supposed to influence" \cite{siegel2024ai}. Metrics such as accuracy reports, such as "How often is the model correct?" for both positive and negative cases, make a bad model look good. For example, if only 1 percent of the customers buy, a model that predicts "no" for every customer achieves 99\% accuracy but fails to predict a potential customer correctly.

\subsection{Evaluating Machine Learning solutions beyond performance}
While there is a broad spectrum of ML techniques for task automation, not all techniques are economically viable in all the scenarios, considering the size of training data and the investment in effort. When in doubt, simple ML techniques are recommended over complex ones \cite{domingos2012few}, \cite{bird2015art}. However, what if a method existed to estimate such investments and compute the benefits when other complex algorithms are chosen? Interestingly, scholarly research exploring the ROI of ML is scarce \cite{nagrecha2016quantifying} \cite{ferrari2005roi}. However, a few discussions and approaches have been proposed in the gray literature, especially on the business front, to weigh the benefits of AI/ML-based software solutions in terms of economic value (dollars). For instance, \cite{altexsoft} emphasizes the need for ROI as a measure for evaluating the benefits of ML.
On the other hand, in another blog \cite{towardsdatascience}, a method is proposed to arrive at ROI economic value, which emphasizes the cost incurred due to prediction outcomes of the ML technique. Although these approaches are intriguing and align well with our agenda, they need more empirical rigour and evaluation. They do not consider the cost of data accumulation and pre-processing, a large part of applied ML effort \cite{figalist2020end}.

Hence, ML techniques must be evaluated beyond performance measures for effective decision-making. In other words, it is necessary to understand how far one should chase the accuracy parameter when the cost of achieving the differential is substantial.

\section{The AROhI Tool}
The AROhI (Data \textbf{A}nalytics and its \textbf{R}eturn \textbf{O}n h\textbf{I}nvestment)\footnote{AROHI also means something which/who is always striving for something more in Sanskrit language}  tool provides users with a cloud-deployed dashboard, enabling estimation of the ROI of Data Analytics. This tool is based on our prior research and experiments \cite{deshpande2020beyond} \cite{deshpande2021bert} \cite{deshpande2021much}.
\textbf{This no-code tool} enables utilizing ML in technology investment-based decision-making through advanced data analysis and ROI estimations, thus empowering non-programmer decision-makers to create and customize ML-driven analytics. The detailed working of the tool has been captured in a demonstration video through the anonymous link to folder\footnote{\href{https://anonymous.4open.science/r/ROI\_tool-1DC8/README.md}{https://anonymous.4open.science/r/ROI\_tool-1DC8/README.md}}. This tool design is also inspired by other tools in the AI realm such as by Labba et al. \cite{labba2023iarch}

\begin{figure*}[!t]
    \centering     \includegraphics[scale=.25, trim={0 2.5cm 0 0}, clip ]{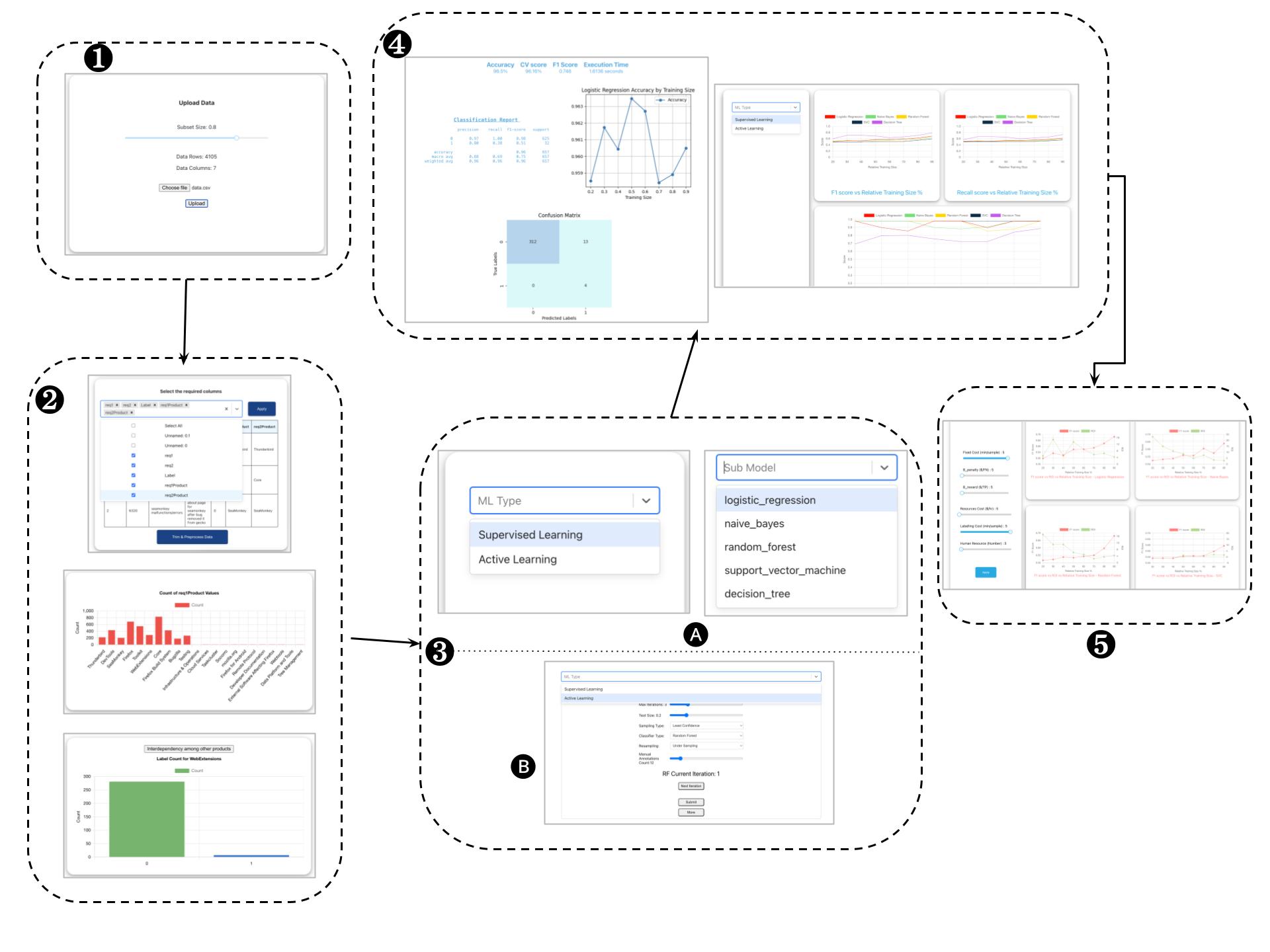}
    \caption{AROhI Tool's user Interface and various screens of the tool enable users to interact and draw conclusions}
    \label{AROhI Interface}
\end{figure*}

\noindent \textbf{Architecture: }Figure \ref{fig: Web Tool Framework} shows the conceptual design of this tool with various technology components used in its development, and figure \ref{AROhI Interface} shows its user interface and capabilities. AROhI has a front-end user interface that was developed using the ReactJs framework. The interactive interface allows end users to upload the dataset and provide additional settings on algorithm selection and cost factors for ROI computation. REST APIs communicate with the backend server developed in the Flask framework. Communication between front end (UI) and back end are managed through REST APIs. The backend server was developed using Python programming language, utilizing the Amazon S3 database for data storage. ML algorithms are developed using SciKit Learn, an open-source software library. 

\noindent \textbf{Machine Learning Algorithms: } 
AROhI comprises supervised learning algorithms, including Logistic Regression, naive Bayes, Decision Trees, Support Vector Classifier (SVC), Random Forest Classifier, and fine-tuned BERT.
The tool also provides an option to use semi-supervised learning, such as Active learning. The user can modify several settings of the active learning algorithm, including the threshold, maximum iteration, test size, sampling type, classifier type, resampling, and manual annotations count. Active learning \cite{settles2009active} attempts to maximize a model's performance gain while annotating the fewest samples possible \cite{ren2021survey}.
With increasing data, the demand and computational cost of data analysis using ML algorithms have gradually risen. Hence, considering both accuracy and investments, it is important to consider a holistic approach to evaluate these algorithms. This approach considers broader business impacts, thus offering a more thorough evaluation of costs and benefits for data analysis.

\noindent \textbf{Features and Capabilities:} Descriptive analysis of the input dataset is shown in the front end (UI) utilizing various charts and graphs (visualization: discussed in use case Section \ref{use case}). The data set is further preprocessed through a data pipeline. Upon uploading the data, the data analysis component (\ding{203} of Figure \ref{AROhI Interface}) enables the end user to observe the features present within the data and further drill down through interactive charts to visualize statistical information through bar graphs. Users can further choose the desired dependent and independent variables (features).
\begin{figure}[!t]
    \centering
    \includegraphics[scale=.37, ]{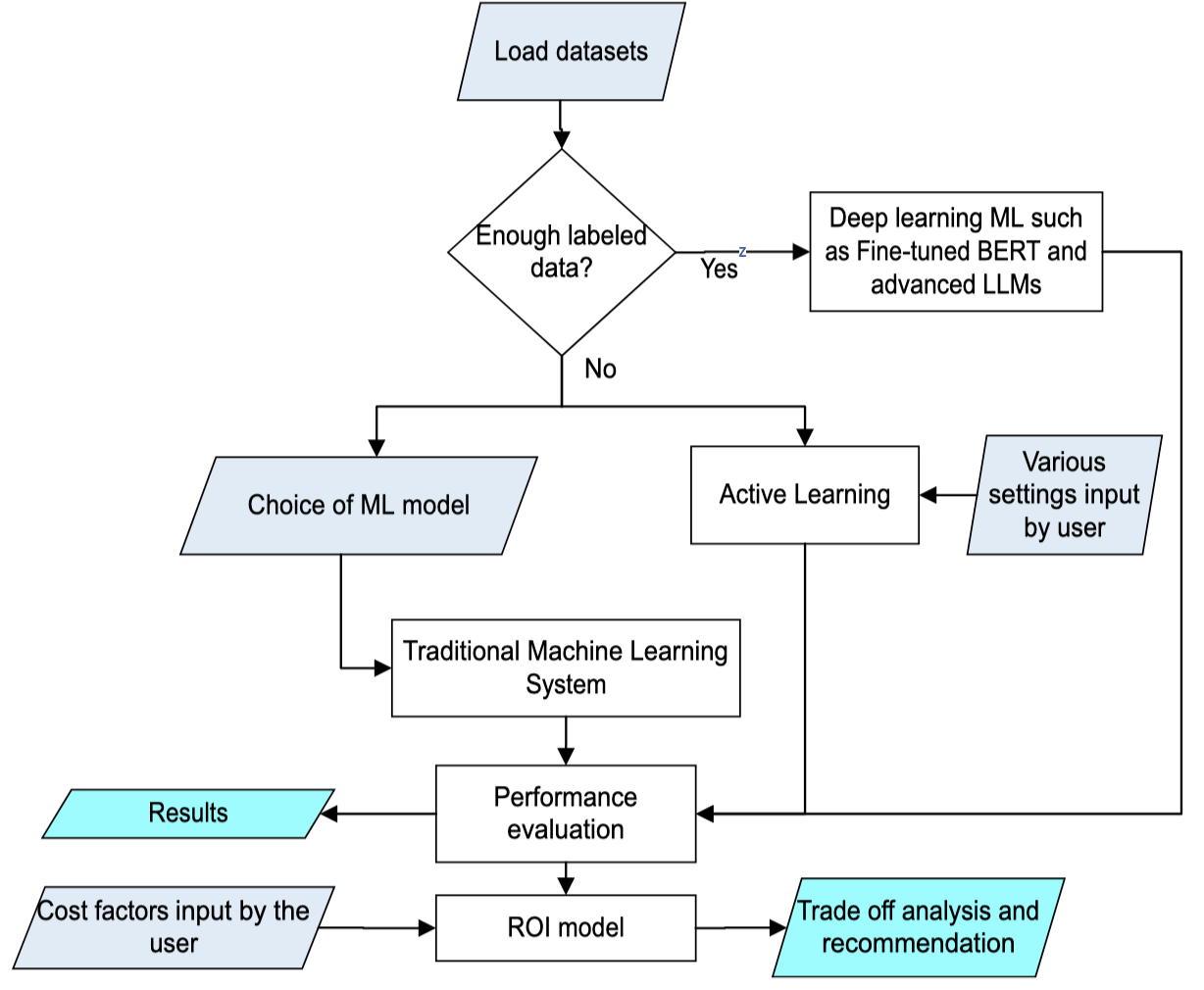}
    \caption{Overall workflow diagram of AROhI tool: Enables end user to choose the desired supervised or semi-supervised method to evaluate its usefulness for their problem and ROI analysis}
    \label{FlowDiagram}
\end{figure}

As shown in the flow diagram of in Figure \ref{FlowDiagram}, AROhI provides an option for the end user to procure additional annotations through employing semi-supervised ML such as Active Learning \cite{settles2009active} if the dataset is smaller or utilizing advanced ML methods such as Fine-tuned BERT (currently implementation is work in progress) over conventional ML algorithms (discussed in this work).

Furthermore, the dataset is processed through a natural language processing (NLP) pipeline. This pipeline includes lemmatization (a process that uses vocabulary and morphological analysis of words to remove inflectional endings and return words to their dictionary form) \cite{zhao2022classification}, and vectorization (a technique in NLP to map vocabulary words to a vector of real numbers.) as part of the text processing. 

The processed data goes to the ML Algorithms during the next stage. Training and testing subsets are created based on the training size provided by the user. We have designed two learning techniques, including supervised and semi-supervised algorithms. 
In supervised learning, annotated data is used to predict the values of the output \cite{hastie2009overview}\cite{reddy2018semi}. Most of the time, the use cases need more sufficient labelled data. Often, labelling data is expensive where unlabeled data is available in abundance \cite{reddy2018semi}. Semi-supervised
learning addresses these issues and works better than supervised learning\cite {reddy2018semi}.

ML models are evaluated based on their accuracy, 10-fold cross-validation score, confusion matrix and F1 score. Additionally, precision and recall are also shown using incremental train set size. Beginning with the 20\% of the training set, models are developed to capture their performance by incrementally growing (increasing the train by 10\%) training set size to capture its impact
on F1 accuracy and ROI. Users can monitor variations in accuracy when training sizes are changed, as the classification matrices are provided to help users better understand their model. The user can also compare all the models based on F1 score, recall and precision in the user interface.

For ROI computations, end users are provided with slider bars to choose various cost factors (discussed in Section \ref{sec:ROI_Modelling}) so that the incremental (with increasing train set) ROI estimates corresponding to the performance of the chosen models are determined. Using sliders, interaction with cost factors configuration generates dynamic graphs for comparative ROI and F1 score analysis for varying training sizes based on user input. The generated graphs are visualized in the user interface as shown in Figure \ref{fig: Web Tool Framework}.

The relationship between the framework of AROhI in Figure \ref{fig: Web Tool Framework} and the AROhI software (tool) in Figure \ref{AROhI Interface} is as follows:
\begin{itemize}
    \item In Figure \ref{fig: Web Tool Framework}, the first component \ding{202} corresponds the figure \ref{AROhI Interface}, depicting data import. Next, the data analysis is performed in component \ding{203} of Figure \ref{AROhI Interface}.

    \item Through REST APIs, the front end communicates with the backend shown in component \ding{203} of the figure \ref{fig: Web Tool Framework}. Data passes through the NLP pipeline, leading to the process in component \ding{204} of Figure \ref{AROhI Interface}, where ML model selection is made. A comparison of various models built using scikit-learn is shown in component \ding{205}.

    \item In the last component of Figure \ref{fig: Web Tool Framework} where ROI computations and configurations of costs happen corresponds to the component \ding{206} of Figure \ref{AROhI Interface}. Finally, the results are sent to the front end (UI) using REST APIs.
\end{itemize}
\begin{table}[!htpb]
    \renewcommand{\arraystretch}{1.3}
    \caption{Parameters used for ROI computation}
    \label{tab:parameters}
    \centering
    \begin{tabular}{p{1cm}|p{3cm}|>{\raggedright\arraybackslash}p{2cm}|c}
 
    & \textbf{Parameters} & \textbf{Symbol} & \textbf{Unit} \\
    \hline
    Cost & Data gathering time & \( C_{\mathrm{dg}} \)& Minutes\\
    & Pre-processing time & \( C_{\mathrm{pp}} \) & Minutes\\
    & Evaluation time & \( C_{\mathrm{e}} \) &Minutes\\
    & Labeling time & \( C_{\mathrm{l}} \) & Minutes\\
    & Resource Cost & \( C_{\mathrm{resource}} \) & \$ per hour\\
    \hline
    Benefit & Value per True Positive instance & \( B_{\mathrm{reward}} \) & \$ \\
    & Penalty per False Negative instance & \( B_{\mathrm{penalty}} \) & \$ \\
    \hline
    Other & H & Human resources & Number \\
    & \( N_{\mathrm{train}} \) & Size of the training set & Number\\
    & \( N_{\mathrm{test}} \) & Size of the testing set & Number \\
    & N & \( N_{\mathrm{train}} \) + \( N_{\mathrm{test}} \) & Number\\
 
    \end{tabular}
\end{table}
\noindent \textbf{ROI Modeling: }\label{sec:ROI_Modelling}
Following a typical approach to the ML life cycle, Planning, Data Preparation,
Execution and Validation are determined as various phases to further associate cost factors with each one as shown in
Table \ref{tab:parameters}. Table \ref{tab:costs} depicts the parameter settings used. These parameters and settings are derived from the work by \cite{deshpande2020beyond}.
Users can configure fixed costs, labelling costs, resource costs, reward per true positive instance, penalty per false negative instance, and the number of human resources used. The fixed cost and labelling cost together form the data preprocessing cost \cite{deshpande2020beyond}. 

\begin{table}[!t]
    \renewcommand{\arraystretch}{1.3}
    \caption{Parameter settings for Binary Requirements Dependency Classification}
    \label{tab:costs}
    \centering
    \begin{tabular}{>{\raggedright\arraybackslash}p{3.5cm}|>{\raggedright\arraybackslash}p{2cm}}
    
    \textbf{Parameters} & \textbf{Value} \\
    \hline
 
    $ C_{\mathrm{fixed}} = C_{\mathrm{dg}} + C_{\mathrm{pp}} + C_{\mathrm{e}} $ & 1 min/sample \\
    \( C_{\mathrm{l}} \) & 0.5 min/sample\\
    \( C_{\mathrm{resource}} \) & \$400/hr\\
    \hline
    \( B_{\mathrm{reward}} \) & \$500/TP\\
    \( B_{\mathrm{penalty}} \) & \$500/FN\\
    \hline
    H  & 1 \\
    N & 4105 \\
     
    \end{tabular}
\end{table}
To demonstrate cost-effectiveness, we focused on financial metrics such as ROI, which is similar to the approach used to evaluate the profitability of software based on real-world cases \cite{stradowski2024costs}\cite{deshpande2020beyond}.

The ROI formula is given by:
\begin{equation}
\refstepcounter{equation} \tag{\theequation}
\text{ROI} = \frac{\text{Benefit} - \text{Cost}}{\text{Cost}} \label{eq:roi}
\end{equation}

The benefit calculation formula is given by:
\begin{equation}
\refstepcounter{equation} \tag{\theequation}
\text{Benefit} = \text{TP} \times B_{\mathrm{reward}} - \text{FN} \times B_{\mathrm{penalty}}
\label{eq:benefit}
\end{equation}

The cost calculations are made by using:
\begin{equation}
\refstepcounter{equation} \tag{\theequation}
\text{Cost} = N\% \times \frac{(C_{\text{fixed}} + C_{\text{l}})} {60} \times H \times C_{\text{resource}}
\label{eq:cost}
\end{equation}
Unit for benefit and cost is the human effort in person-hour\cite{deshpande2021bert}.

\subsection{Evaluation Metrics}
We are letting users evaluate their results using the following metrics:

\begin{itemize}
    \item \textbf{Cross validation: } It is statistical method of evaluating
    Furthermore, comparing learning algorithms by dividing data into two segments: one used to learn or train a model and the other used to validate the model \cite{refaeilzadeh2009cross}. We have used k-fold cross-validation with k = 10.
    \item \textbf{Confusion Metrics:} We obtained a $ 2 \times 2 $ matrix for each model indicating Actual Negative, Actual Positive, Predicted Negative and Predicted Positive \cite{deshpande2021bert}.
    
    \item \textbf{F1 Score:} It is calculated using actual and predicted classes.
    \begin{equation}
    \refstepcounter{equation} \tag{\theequation}
    \text{F1 score} = \frac{( 2 \times TP)}{2 \times TP + FP + FN}
    \label{eq:f1score}
    \end{equation}

    \item \textbf{ROI:} As shown in Equation \eqref{eq:roi}, it is a financial metric used to evaluate the profitability of an investment.
\end{itemize}

\section{Use case}
\label{use case}
\textbf{To demonstrate the utility of this tool,} we utilize the Requirements Dependency Extraction problem for the dataset from the Mozilla family of products \cite{mozilla2020bugzilla} as a use case (such as Firefox browser, Thunderbird email client, etc.). A requirement is a condition or capability a user needs to solve a problem or achieve an objective. It is also a condition or capability that must be met or possessed by a system or system component to satisfy a contract, standard, specification, or other formally imposed documents \cite{7435207}. To demonstrate the utility of our solution, we performed binary classification of requirements dependency types using a subset of the entire dataset for training and testing purposes. This process was iterated by gradually increasing the proportion of the training set and storing the results for further calculations. \textbf{This allows users to }understand the trade-off between the F1 score and ROI and the eventual break-even point of positive returns.
\begin{table*}[!htpb]
\centering
\renewcommand{\arraystretch}{1.3}
\caption{Samples of dependency pairs from our use case dataset.}    
\begin{tabular}{p{2.7cm}|p{6cm}|p{6cm}}
\textbf{Dependency type} &  \textbf{Requirement 1}                                                                     &  \textbf{Requirement 2}                                                                                   \\ \hline
\multirow{4}{*}{\textit{\small {DEPENDENT}}}      & implement quantum bar design update behind pref &  shift bookmark toolbar downwards when urlbar is focused in new tab                                                       \\ \cline{2-3}
 & update navigation toolbar overflow panel for photon                                      &  make bookmark toolbar item blend in better in overflow panel                                                      \\
\hline \hline
\multirow{4}{*}{\textit{\small{INDEPENDENT}}} &  offer choice to open natively with content disposition attachment view/show in browser                                & remove old onboarding add on prefs            \\ \cline{2-3}
& context menu option to open link/image using external application &  add automated test for item from bookmark toolbar can be sorted by name\\

\end{tabular}
 \label{tab:dependency}
\end{table*}
\subsection{Dataset}
Our work adopts the Binary Requirements Dependency Classification concept presented by \cite{deshpande2021much} and \cite{deshpande2021bert}. Given a set of requirements R and a pair of requirements \( (r, s) \in R \times R \), we have considered two requirements as INDEPENDENT if addressing one of them does not have any logical or practical implications for addressing the other. Otherwise, they are labelled as DEPENDENT\cite{deshpande2021much}. 
Table \ref{tab:dependency} represents pairs of requirements dependencies from our dataset. 
\subsection{Execution}
\begin{itemize}
    \item To select the initial subset of the data, we used its 80\%. We kept 20\% for testing and validation from the subset of training data depicted in \ding{202} of Figure \ref{AROhI Interface}.
    
    \item component \ding{203} of Figure \ref{AROhI Interface} shows that requirements and their descriptions were selected during feature selection. Figure \ref{fig: Web Tool Framework} shows the tool's workflow when the user inputs the data through front end (UI). However, users can customize the selection using the dropdown in the tool. The chart is displayed on the front end to determine the characteristics and quantify the number of various requirements in the data.
    
    \item In the next component, the user can compare the model results for semi-supervised shown in \tikz[baseline=(char.base)]{    \node[shape=circle,draw,fill=black,text=white,inner sep=0pt, minimum size=5pt] (char) {\scriptsize 3b};
    } and supervised learning in \tikz[baseline=(char.base)]{
    \node[shape=circle,draw,fill=black,text=white,inner sep=0pt, minimum size=5pt] (char) {\scriptsize 3a}; in Figure \ref{AROhI Interface}
    } using different classifiers.
    
    \item In Active Learning, the user can choose from available sampling techniques, threshold, resampling technique, manual annotation count and classifier type. Component \ding{204} of Figure \ref{AROhI Interface} illustrates, how the ML component of the tool works. This component stores the true positive and false negative count. Next is to compare different models regarding F1 score, recall and precision shown in component \ding{205}, Figure \ref{AROhI Interface}. The final component in \ding{206} is to calculate benefit and cost using the data the user selects through slider and ML models stored data.
\end{itemize}

\begin{table*}[!htpb]
\renewcommand{\arraystretch}{1.2}
\caption{F1 and ROI for the settings in Table \ref{tab:costs}}
\centering
\begin{tabular}{l|ll||ll||ll||ll||ll}
 & \multicolumn{2}{l||}{\textbf{Logistic   regression}} & \multicolumn{2}{c||}{\textbf{Naïve Bayes}} & \multicolumn{2}{c||}{\textbf{Random Forest}} & \multicolumn{2}{c||}{\textbf{SVM}} & \multicolumn{2}{c}{\textbf{Decision Tree}} \\  
\textbf{Training size} & \multicolumn{1}{l|}{\textbf{F1}} & \textbf{ROI} & \multicolumn{1}{l|}{\textbf{F1}} & \textbf{ROI} & \multicolumn{1}{l|}{\textbf{F1}} & \textbf{ROI} & \multicolumn{1}{l|}{\textbf{F1}} & \textbf{ROI} & \multicolumn{1}{l|}{\textbf{F1}} & \textbf{ROI} \\ \hline
20\% & \multicolumn{1}{l|}{0.50} & 311.50 & \multicolumn{1}{l|}{0.52} & \textbf{1249.00} & \multicolumn{1}{l|}{0.52} & 936.50 & \multicolumn{1}{l|}{0.49} & -1.00 & \multicolumn{1}{l|}{0.60} & -938.50 \\  
30\% & \multicolumn{1}{l|}{0.54} & 832.33 & \multicolumn{1}{l|}{0.53} & 832.33 & \multicolumn{1}{l|}{0.51} & 415.67 & \multicolumn{1}{l|}{0.49} & -1.00 & \multicolumn{1}{l|}{\textbf{0.72}} & \textbf{2082.33} \\  
40\% & \multicolumn{1}{l|}{0.53} & 311.50 & \multicolumn{1}{l|}{0.54} & 624.00 & \multicolumn{1}{l|}{0.54} & \textbf{467.75} & \multicolumn{1}{l|}{0.49} & -1.00 & \multicolumn{1}{l|}{0.70} & 1717.75 \\  
50\% & \multicolumn{1}{l|}{0.57} & \textbf{749.00} & \multicolumn{1}{l|}{0.56} & 499.00 & \multicolumn{1}{l|}{0.53} & \textbf{374.00} & \multicolumn{1}{l|}{0.50} & \textbf{124.00} & \multicolumn{1}{l|}{0.69} & \textbf{749.00} \\  
60\% & \multicolumn{1}{l|}{0.58} & 519.83 & \multicolumn{1}{l|}{0.56} & 311.50 & \multicolumn{1}{l|}{0.54} & \textbf{311.50} & \multicolumn{1}{l|}{\textbf{0.51}} & 103.17 & \multicolumn{1}{l|}{0.64} & -1.00 \\  

70\% & \multicolumn{1}{l|}{0.59} & \textbf{356.14} & \multicolumn{1}{l|}{0.60} & 445.43 & \multicolumn{1}{l|}{0.55} & \textbf{266.86} & \multicolumn{1}{l|}{\textbf{0.51}} & \textbf{88.29} & \multicolumn{1}{l|}{\textbf{0.68}} & -90.29 \\ 

80\% & \multicolumn{1}{l|}{0.62} & \textbf{389.63} & \multicolumn{1}{l|}{0.57} & \textbf{233.38} & \multicolumn{1}{l|}{0.60} & \textbf{311.50} & \multicolumn{1}{l|}{0.55} & \textbf{155.25} & \multicolumn{1}{l|}{\textbf{0.68}} & \textbf{311.50} \\ 

90\% & \multicolumn{1}{l|}{\textbf{0.68}} & \textbf{276.78} & \multicolumn{1}{l|}{\textbf{0.64}} & \textbf{207.33} & \multicolumn{1}{l|}{\textbf{0.68}} & \textbf{276.78} & \multicolumn{1}{l|}{\textbf{0.59}} & \textbf{137.89} & \multicolumn{1}{l|}{\textbf{0.72}} & \textbf{207.33} \\
  
\end{tabular}
\label{ROIvsF1}
\end{table*}
\subsection{Results}
Table \ref{tab:accuracy} shows the accuracies of various supervised learning models where execution time denotes the time taken to execute the process from triggering the model for training for the initial set training size defined by the user (0.8 in this case). We also performed the 10-fold cross-validation during training of this initially set data. Table \ref{ROIvsF1} shows the variation of F1 score and ROI with training data size. 

For Logistic Regression: The best accuracy can be reached with a dataset of up to 60\%. Beyond that, the ROI drops. Although the highest accuracy is with 90\% of the dataset, the ROI drops sharply with the given settings: 

Naive Bayes: The highest ROI is with just 20\% of the dataset, so investing more is not useful. although NB's performance becomes comparable with others at a certain point, it does not yield value beyond 40\% at .54 accuracy. Thus, NB might not be a good algo for the given problem.

Random Forest: Accuracy does not grow until 80\% of the train set however, similar to NB, ROI is maximum at just 20\% of the dataset, so even RF might not be a good algorithm to choose from.

SVC: Although ROI is positive only after 50\% of the dataset is procured for training. Also, the accuracy is poor, and the algorithm struggles to understand the dataset's underlying pattern, making this one an inferior choice. Also, beyond 80\%, the ROI drastically decreases while accuracy increases marginally. Also, it shows that the break-even is achieved only after investing in 40\%  training data acquisition.

Decision Tree:  Achieves best accuracy of .72 just with 20\% of the dataset with the highest ROI, making this one the most valuable and effective algorithms for the given problem and values.

\subsection{ROI Sensitivity Analysis}
\begin{figure*}[!ht]
\begin{minipage}{.45\textwidth}
    \centering
\includegraphics[scale=.18, trim={0 1.5cm 0 1cm}, clip]{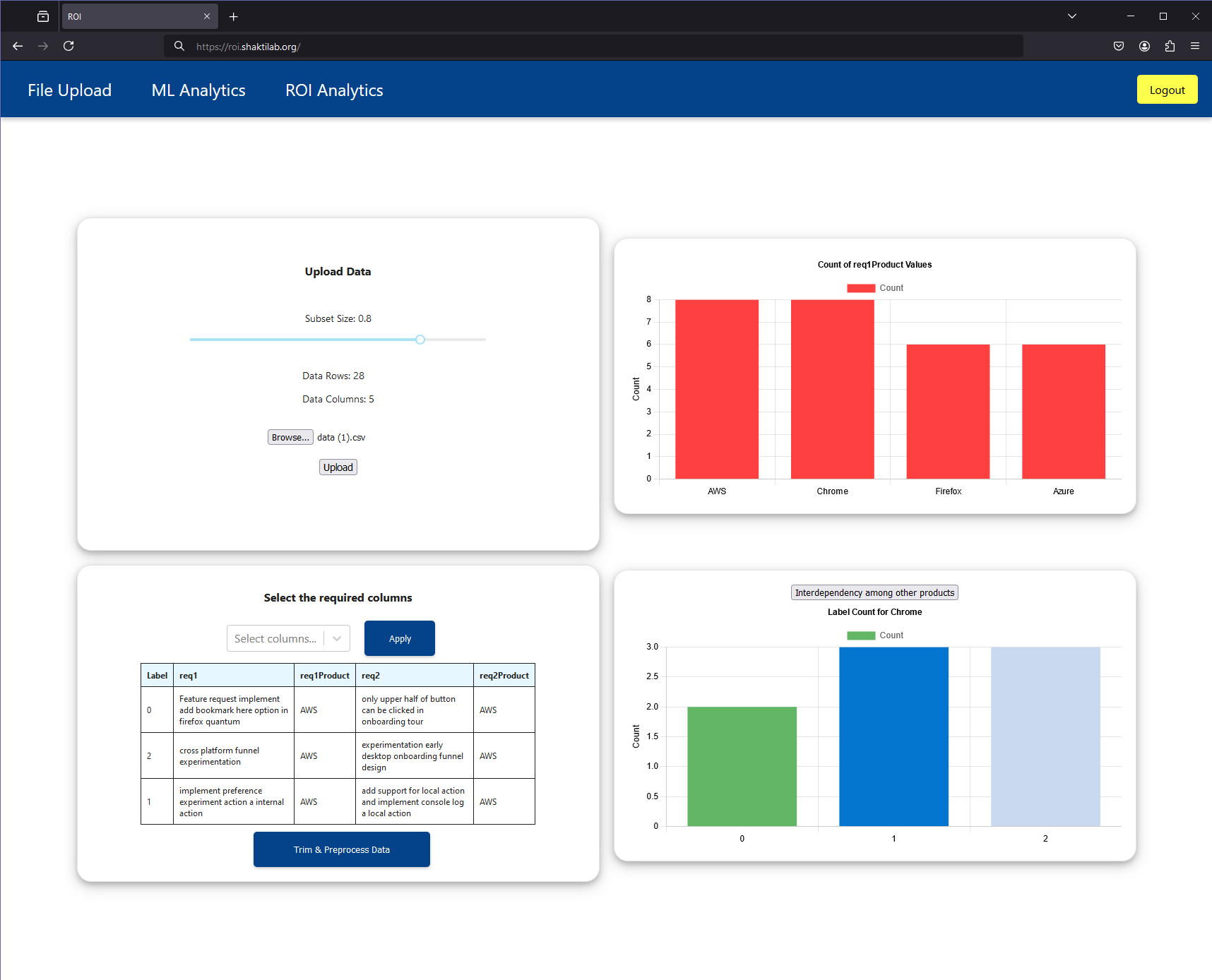}
    \caption{The login based UI shows descriptive analysis of the uploaded data}
    
    \label{fig:vis1}
    \end{minipage}
    \hspace{10mm}
    \begin{minipage}{.45\textwidth}
    \centering
   \includegraphics[scale=.18, trim={0 1.5cm 0 1cm}, clip]{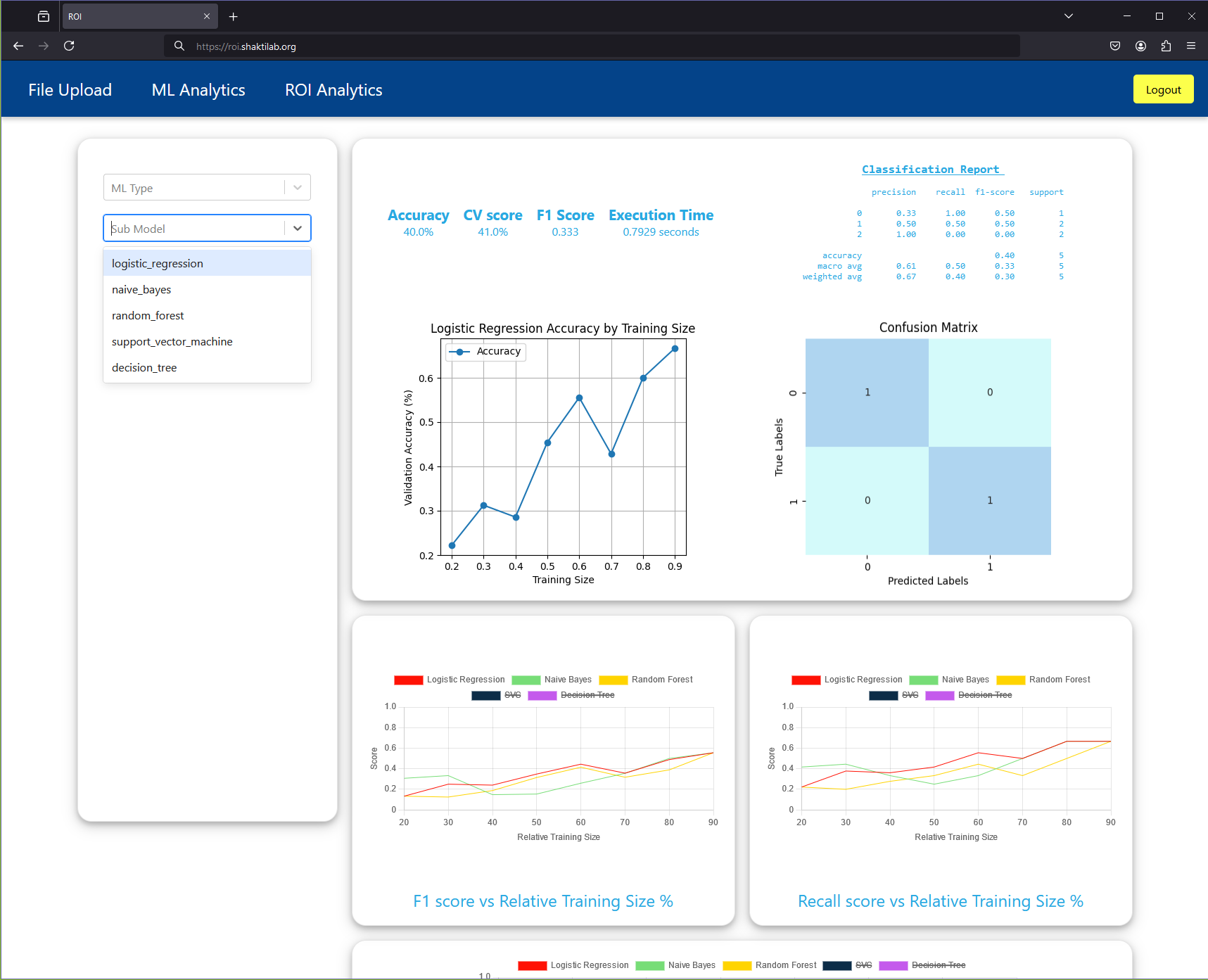}
    \caption{Various graphs showing relative train set and various ML measures such as F1-Score, Precision and Recall}
    \label{fig:vis2}
    \end{minipage}
\end{figure*}

\begin{figure*}[!ht]
\begin{minipage}{.45\textwidth}
    \centering
\includegraphics[scale=.18, trim={0 1.5cm 0 1cm}, clip]{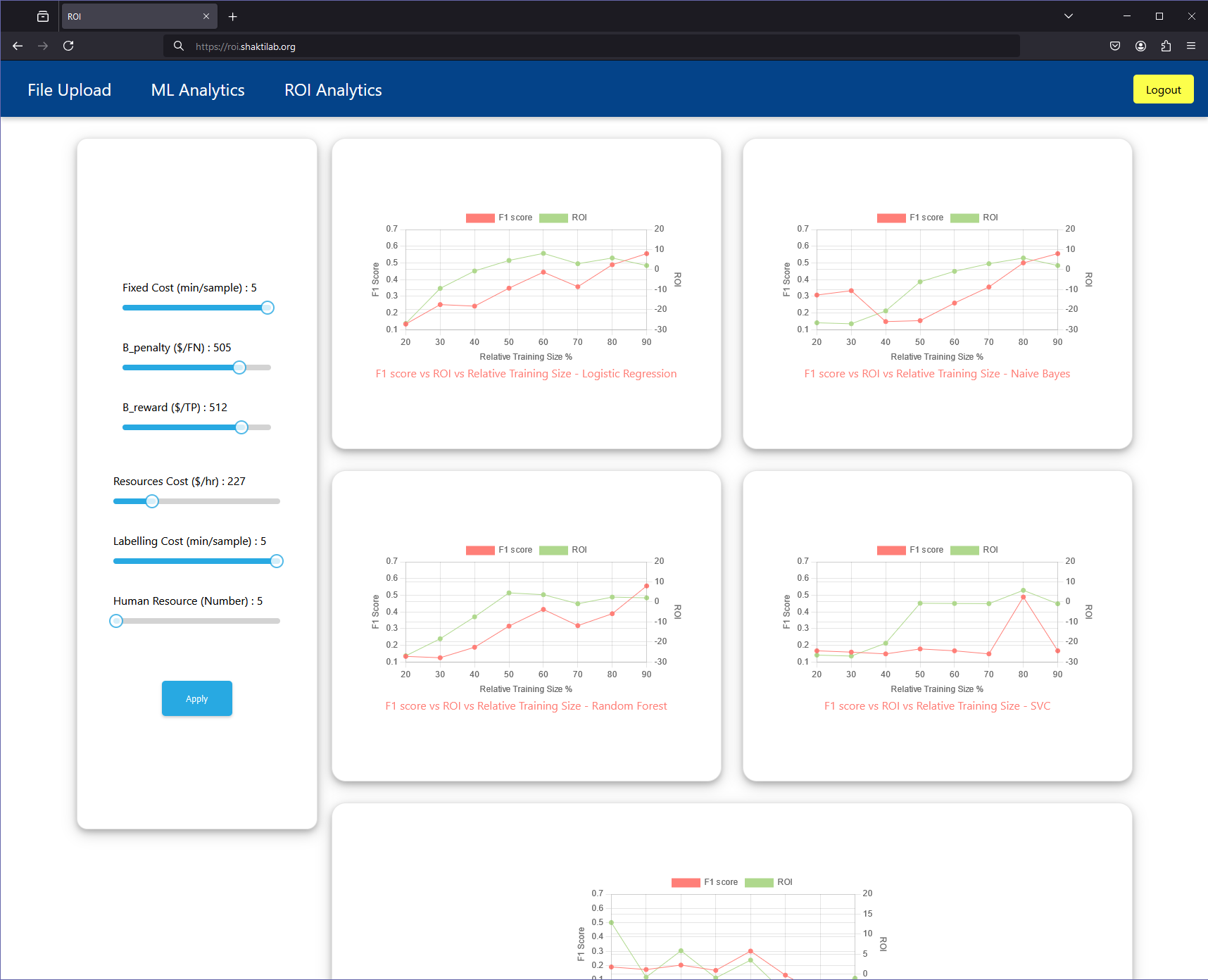}
    \caption{For the chosen cost factors, ROI vs F1 score charts are generated dynamically showing at what point break even is achieved and how the ROI varies with increasing train set for ML algorithms}
 \label{fig:vis3}
   
    \end{minipage}
    \hspace{10mm}
    \begin{minipage}{.45\textwidth}
    \centering
   \includegraphics[scale=.18, trim={0 1.5cm 0 1cm}, clip]{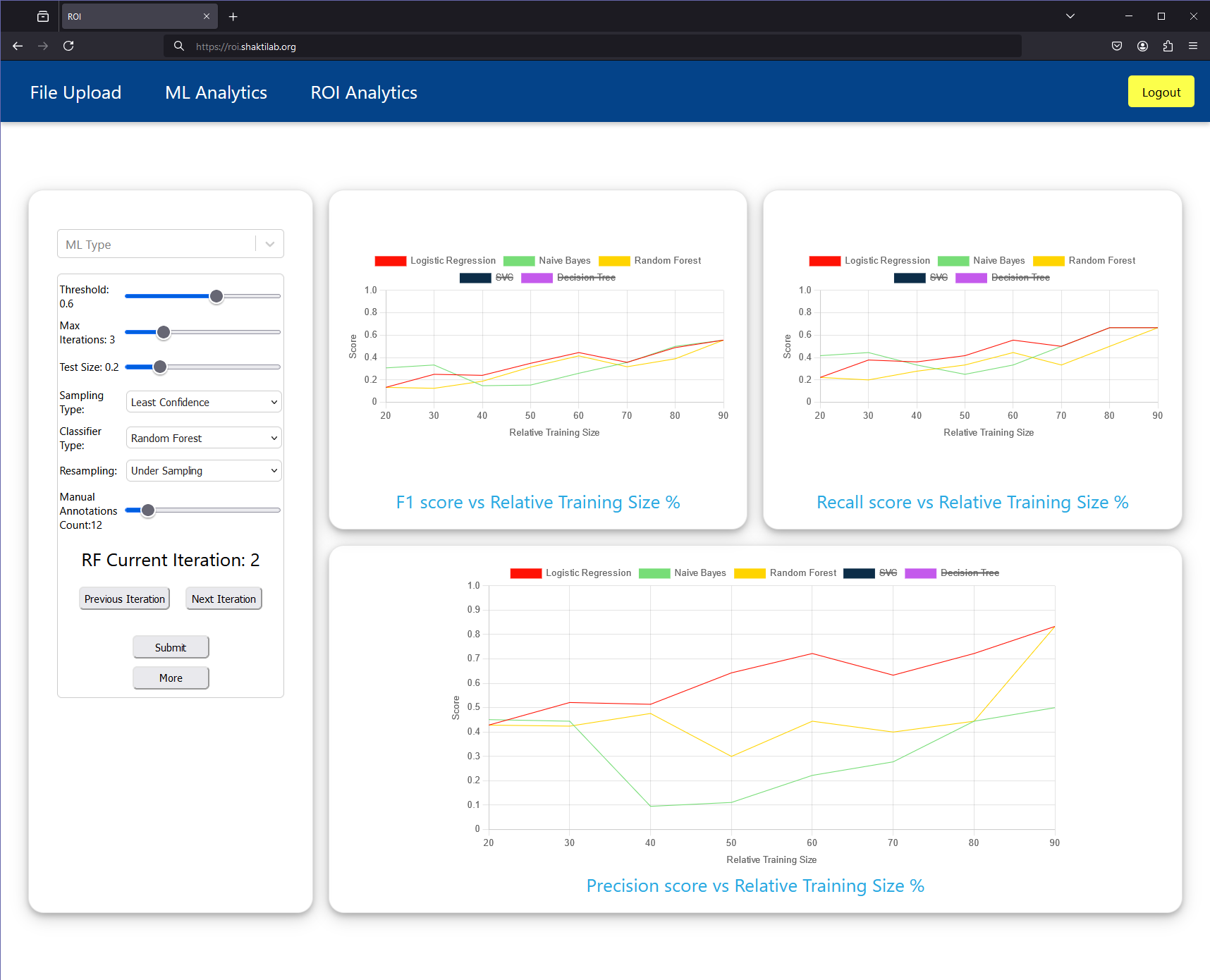}
    \caption{For various settings using the sliders on the left side, Active Learning based algorithms can be used to accumulate additional annotations incrementally. This is work in progress}
    \label{fig:vis4}
    \end{minipage}
\end{figure*}

As an outlook to varying input costs, we envision enhancing the tool further to show how ROI varies by varying the various cost factors from Table \ref{tab:costs}.  
For example, we fixed the training size to 80\% for Logistic Regression under supervised learning and then increased Resource Cost from \$400 to \$440 (10\% increase), and ROI decreased from 389.62 to 354.11 ( 9.11\% drop). A detailed analysis has been explained and shown in our past work \cite{deshpande2021bert} through heat map visualizations. Enabling such visualization remains part of the AROhI tool's feature list. 
As show in Figures \ref{fig:vis1}, \ref{fig:vis2}, \ref{fig:vis3}, and \ref{fig:vis4} to visualize such analysis dynamically.

\begin{table}[!t]
    \renewcommand{\arraystretch}{1.4}
    \caption{Accuracy of Supervised Learning Models
    fixing train size = 0.8}
    \label{tab:accuracy}
    \centering
    \begin{tabular}{>{\raggedright\arraybackslash}p{2cm}ccc}
    \textbf{Model} & \textbf{Accuracy \%} &  \textbf{CV score \%} &
     \textbf{\shortstack{Execution Time\\(sec)}} \\
     \hline
    Logistic Regression & 96.50 & 96.16 & 1.6053 \\
    \hline
    Decision Tree Classifier &  94.82 & 95.95 & 2.3645\\
    \hline
    Naive Bayes & 92.39 & 96.07 & 0.781\\
    \hline
    SVC & 95.89 & 96.07 & 9.3548\\
    \hline
    Random Forest Classifier & 95.28 & 96.16& 21.7949 \\
   
    \end{tabular}
\end{table}

\section{Limitations}
This research has several limitations: (i) The use case evaluation is restricted to a small dataset that uses less intense computational resources. Rigorous testing of the tool remains part of our future work. However, based on our prior work, we are confident the results are generalizable. (ii) The tool currently supports classification problems; Extending to other problem types remains out of scope. (iii) The tool is trained on independent and dependent types of requirements, which we plan to extend to other multi-class classification problems in the future. (iv) Other aspects of the ML, such as bias, fairness, and carbon footprint, have implications for ROI. Thus, instead of evaluating the models based on performance measures and ROI alone, we consider these aspects while computing ROI. However, this remains out of scope at this point.

\section{Conclusion and Future Work}
Based on our prior research work, in this study, we propose the AROhI tool to compute the ROI of data analytics and compare it with the F1-score, thus enabling end users (stakeholders and decision-makers). This tool provides an extensive and interactive interface to analyze and compare various data analysis techniques using ML based on accuracy, costs and benefits, thus providing a holistic ROI-driven approach to any classification problem. In our future work, we aim to overcome the current limitations of our tool and improve its performance and applicability across a broader range of datasets and ML as well as LLM models. Further, we plan to address the following areas:
\begin{enumerate}
    \item We intend to generalize our approach to all dependency classification datasets, regardless of the specific column names used within such datasets. This will ensure that our model is versatile and adaptable to different data types. 
    \item In addition, we plan to apply unsupervised learning algorithms, improving the model's ability to detect underlying patterns without relying on labelled data.
    \item Furthermore, we intend to extend our methodology to incorporate advanced language models such as BERT, GPT, and other Large Language Models (LLMs) for ROI calculations.
    \item Along with ROI and accuracy, we intend to introduce a metric to evaluate bias and fairness in the dataset.
\end{enumerate}

\bibliographystyle{ieeetr} 
\bibliography{references}

\end{document}